\documentclass[10pt]{iopart}

\usepackage{iopams}  
\usepackage{setstack}  
\usepackage{graphicx}
\usepackage{epstopdf}
\usepackage[table]{xcolor}

\begin{document}

\title{Scalable, chip-based optically-controlled gates for quantum information processing}

\author{I. A. Burenkov}
\address{Physics Department, Lomonosov Moscow State University, Moscow 119991, Russia}
\author{O. V. Tikhonova}
\address{Physics Department, Lomonosov Moscow State University, Moscow 119991, Russia}
\address{Skobeltsyn Institute of Nuclear Physics, Lomonosov Moscow State University, Moscow, Russia}
\author{S. V. Polyakov}
\address{National Institute of Standards and Technology, Gaithersburg, Maryland 20899, USA}
\ead{sergey.polyakov@nist.gov}


\begin{abstract}
Here we present a simple and robust method to build on-the-fly configurable quantum gates based on a photonic exchange between quantum nodes. The idea is based on a high reflectivity of Bragg grating structures near resonant wavelengths. The control is exerted by applying an external strongly off-resonant or even a static electromagnetic field and taking advantage of the Kerr effect. When the nonlinear phase shift is strong enough, the Bragg mirror disappears, thereby allowing a transmission of a wave packet from one node to another. An example of a protocol for quantum logic gates that relies on this framework is offered.
\end{abstract}

%
\vspace{2pc}
\maketitle
%
%
\ioptwocol

\section{Introduction}

The progress of quantum information science has been very significant. On the theory side, multiple algorithms that take advantage of quantum information processing have been developed \cite{Ekert, Shor, DiVi1}. Multiple physical systems, particularly those based on trapped ions, neutral atoms, and quantum dots \cite{Ions, QDot, RAtoms, Atoms} have been surveyed and found suitable as quantum nodes. To be useful, a physical carrier of quantum information should couple well with the other carriers when needed, but otherwise, must be isolated from other carriers and the environment. In practice, satisfying this condition is hard, and requires a major technical effort. At present, only a few nodes have been demonstrated to be fully controlled simultaneously \cite{Monroe, MIons, spin}. This comes at a cost of a significant experimental overhead, impeding further growth. Yet, to support a practical quantum information algorithm, the number of gates must be large. Thus, an additional criterion arises - the ability to manufacture and link a large number of nodes, and maintain a high degree of control over those nodes with little additional technical overhead to support it - or scalability. The most promising candidates for scalability are physical systems based on solid-state nodes, such as quantum dots and NV centers \cite{QDot, NVC}. Thus, the main problem and the goal of an investigation is to combine the scalability with logic operations ``on demand''.

A basic quantum processing unit consists of nodes (i.e. atoms, quantum dots, etc.) that are separated from the environment (for example, with photonic crystals) and coupled to on-chip waveguides for mutual interaction and an optical readout. To date, much effort is directed towards the development of quantum information processing systems with nodes connected by passive optical interconnects.  A passive optical coupling between quantum nodes simplifies the interconnection but makes the design of nodes more complex. Thus, passive optical interconnects may be more suitable for long-distance quantum communication. For a scalable quantum information transfer between nearby nodes within a quantum processing block (i.e. on a chip), it might be beneficial to use simple two-level systems as nodes and use actively-switched channels. In this letter, we propose quantum processing units that take advantage of actively controlled optical channels based on a natural property of solids - the Kerr nonlinearity. We show that actively-controlled optical channels are a simple, robust, versatile and highly scalable approach to configuring and operating quantum gates.

This manuscript is structured as follows. First, we introduce the idea of a quantum protocol based on stop-band mirrors and cross-phase modulation switching. Next, we discuss an isolated node-cavity system as an elementary block of this scheme. Thirdly, the photon exchange is discussed. Then we show how to use photon exchange and single-qubit inversion to build key two-qubit quantum gates. A feasibility study providing technical details to aid future experiments is offered as a supplementary information.

\section{Overview of the method}

\begin{figure}[t]
\includegraphics[width=0.95\columnwidth]{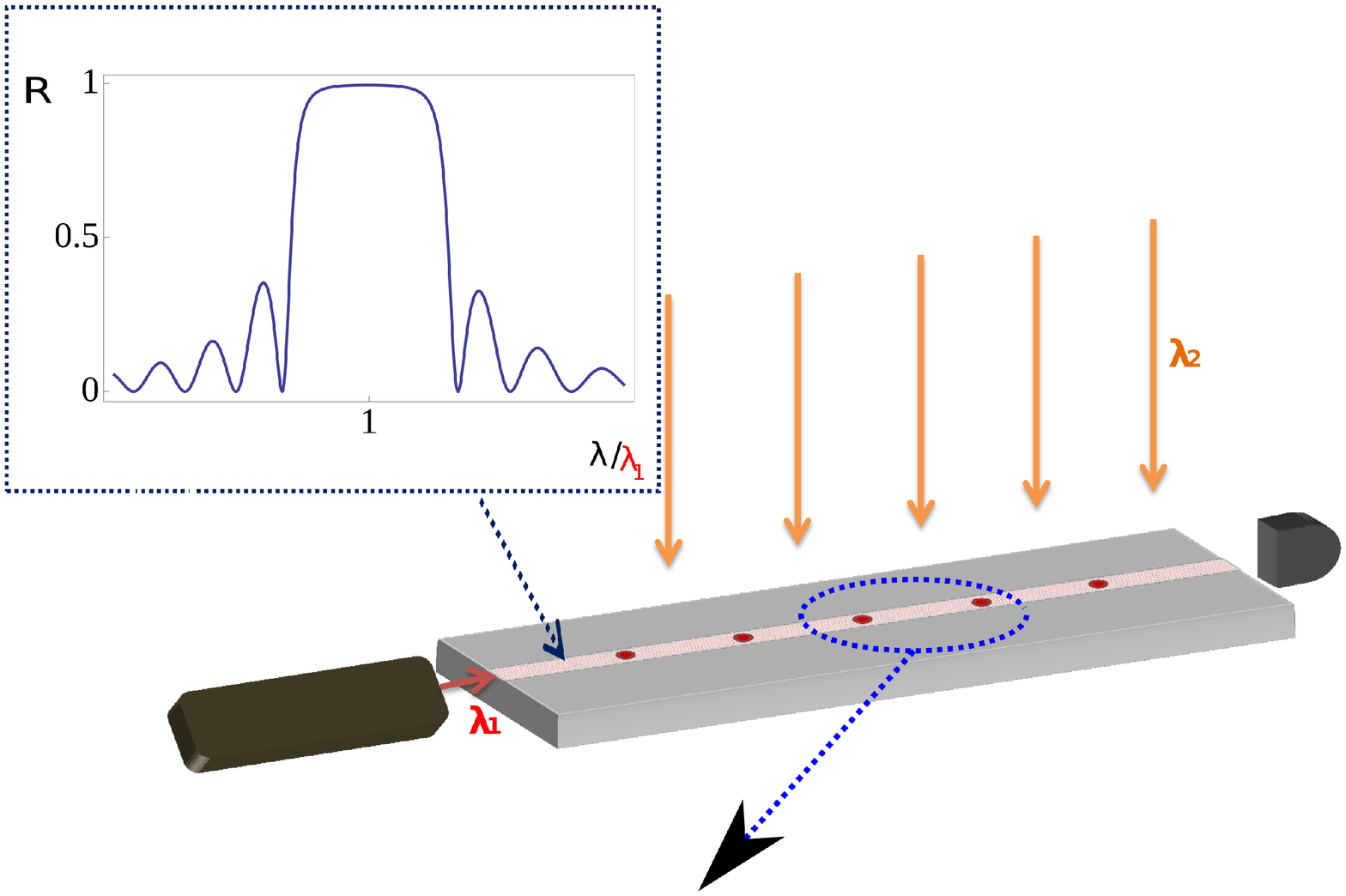}

\includegraphics[width=0.95\columnwidth]{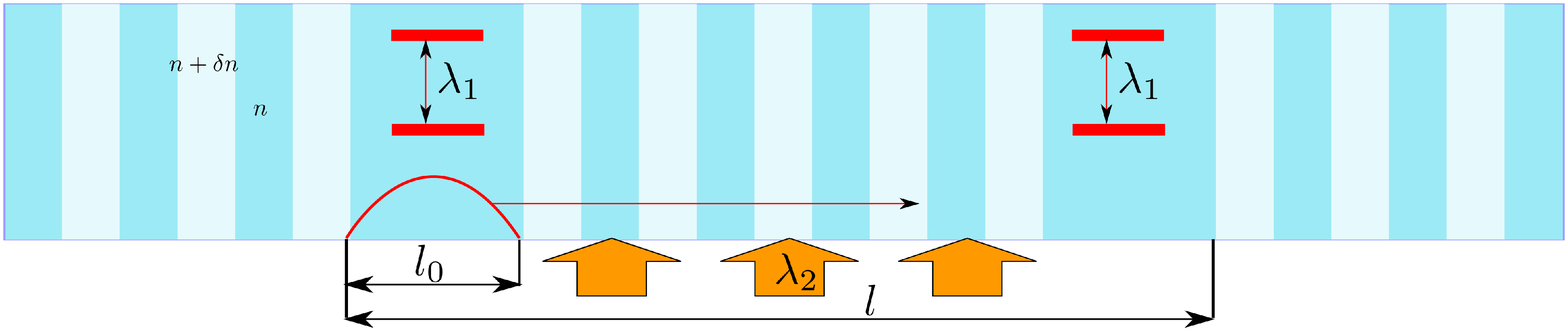}
\caption{\label{fig:one} (Color online) A scalable 1-dimensional quantum register with a nonlinear Bragg waveguide connecting several N-level quantum nodes. At least one optical transition wavelength $\lambda_1$ between node levels $|e\rangle$ and $|g\rangle$ is in the stop band of a Bragg waveguide, turning the channel into a high-reflective cavity for efficient light-matter coupling. The Bragg structure is controlled via a cross-phase modulation due to a Kerr nonlinear response to the external laser beam $\lambda_2$, incident from above the chip. The inset shows the typical reflectivity dispersion curve of the Bragg structure. }
\end{figure}

Our proposed optical channel connects two or more N-level quantum nodes by a nonlinear Bragg waveguide, whose stop-band contains $\lambda_1$, the wavelength of an optical transition of the quantum nodes (see Fig. \ref{fig:one}). Ordinarily, the high reflectivity of a Bragg grating structure (see the inset in Fig. \ref{fig:one}) ensures isolation between the nodes, and helps maintain high-Q coupling due to Bragg reflection \cite{Nozaki}. With the grating in place, the nodes are isolated, and independent manipulation of single nodes becomes possible through deterministic interaction with a classical laser pulse. This enables single-qubit gates. We refer to this configuration as an ``isolated node-cavity system''. At the same time, the Bragg resonance of the nonlinear structure can be removed either with an external far off-resonant optical \cite{Schneider} or a static electrical field \cite{Jin} (shown as a field with a wavelength $\lambda_2$ in Fig. \ref{fig:one}) due to the cross-phase Kerr effect. 
With the Bragg mirror removed, a ``common node cavity'', which connects two or more nodes, occurs. A common node cavity enables photon exchange between the nodes, giving rise to multi-qubit gates.  Because we adopted the classical nonlinear Kerr switching for quantum gates, the range of material systems exhibiting the necessary nonlinearity is very broad. The enabling calculation and experimental results, then, are directly applicable to any of those materials, after scaling by their Kerr coefficient.



The advantage of this scheme is its high degree of versatility in real-time. That is, single-qubit or many-qubit gates are supported, they can be configured on demand and switched on and off as needed. Because a photon exchange between a pair of nodes is sufficient to support most common quantum gates, a discussion of multi-node photon exchange is outside the scope of this manuscript. 

Here we demonstrate our concept for quantum nodes that are single two-level quantum systems, for example, quantum dots or surface-trapped atoms. In this arrangement the system ``atom+field'' is used as a qubit, acting as storage, as single qubit gate and initialization/readout; whereas a pair of nodes offers two-qubit gates. We point out that protocols discussed here require strong coupling of nodes and cavities because we used the most basic, 2-level systems as nodes. The use of N-level systems would eliminate the need for strong coupling (see, for example, \cite{Monroe1} and references therein). An adaptation of our method to N-level systems, however, is beyond the scope of this Letter.


\section{Isolated node-cavity system and one-qubit gates}

Each node is surrounded by Bragg mirrors giving rise to a set of isolated cavities with an effective length $l_0$. A node is coupled to one of the cavity eigenmodes: ${F}_{s_0}(z,l_0)=\sqrt{2/l_0}\sin{(\pi s z/l_0)}$, where $s_0$ is the mode number. 
The isolated node-cavity system with one quantum of excitation undergoes vacuum Rabi oscillations between the exited state of the atom and a single excitation of the cavity mode. The coupling between the node and the cavity is given by the Rabi frequency $\Omega_0$, which depends on a dipole moment $d_{ge}$, a transition wavelength $\lambda_1$ and a mode volume $V$.  The expression for Rabi frequency in SGC is:
\begin{equation}
\Omega_0=d_{if}\sqrt{\frac{8\pi^2 c}{\hbar V \lambda_1}},
\label{eqn:Rabi}
\end{equation}
where $\hbar$ is the Planck constant and $c$ is the speed of light in vacuum. As demonstrated by prior work on quantum dots coupled to Bragg resonators \cite{Pelton, StrongCouple, StrongCouple2}, the mode 
volume can be small, i.e. on the order of $(\lambda/n)^3$, and the dipole moment as large as $\approx10^{-28}\mathrm{C}\cdot\mathrm{m}\approx10$ $\mathrm{ea}_0$ can be achieved, yielding the vacuum Rabi frequency of $\Omega_0\approx 10^8\div10^9$ rad$\cdot$s$^{-1}$. To achieve such a frequency \cite{Bose}, the transparency of Bragg mirrors in the closed channel should be less than $10^{-6}$, which is within experimental reach \cite{Pierno}. 

For a single node with vacuum Rabi oscillations, a natural choice of basis for qubit states are bipartite states of the system ``atom+field'': $|0\rangle=|e\rangle|0\rangle$ and $|1\rangle=|g\rangle|1\rangle$. These states can be read out in a deterministic way by creating a transparency channel between a cavity and a detector with an efficiency comparable to that of a single photon detector (For systems with more than two ``atomic'' levels, the qubit state can be measured differently: via a transition to an axillary exited state of the ``atom'').
An initialization of the nodes can be done by opening a transparency channel between a single photon source and a node, i.e. similarly to read-out. A single photon source can be made with a laser and one dedicated node, through various mechanisms described in the literature. One can use a resonant classical laser $\pi/2$- pulse \cite{EberlyBook} or adiabatic rapid passage \cite{APass} or a photon blockade \cite{PhBlockade}.
 In the first case, the resonant external laser field applied to the system provides a deterministic initialization of the node with $\pi$-pulses, $\pi/2$-pulses, etc. With the proper choice of the intensity of the initialization classical field, the initialization time can be made several orders shorter than one period of vacuum Rabi oscillations of the node. Since initialization is an operation on a single isolated node, multiple nodes can be initialized in parallel. 

Several one-qubit gates can be implemented on isolated nodes. For instance, the ``inversion'' operator can be implemented by freezing Rabi oscillations in selected nodes by sending a set of classical $2\pi$ pulses \cite{EberlyBook} and allowing these nodes to remain in their initial quantum states while letting the other nodes evolve.  
\section{Common node-cavities and photon exchange}

Two-qubit gates are mediated by a high-fidelity photonic exchange between the nodes. Here we describe the use of cross-phase modulation that turns the Bragg mirror that separates the two isolated nodes into a transparent waveguide. Once the control field is applied between a pair of nodes, the mirror disappears and the photon exchange occurs. Because  a typical photon exchange duration is much faster than any interactions between the nodes and cavities, only a photonic part of a bipartite qubit will be affected. While the corresponding photon wavefunction is in an eigenmode of an isolated cavity, it is not in an eigenmode of a much larger common cavity. Then, the propagation of the photon is given by a solution of the wave equation, which has the form:
\begin{equation}
\frac{\partial^2 {E}}{\partial z^2}-\frac{n^2}{c^2}\frac{\partial^2 {E}}{\partial t^2}=0.
\label{eqn:wave}
\end{equation}
where $n$ is the effective refractive index of the waveguide.
Eq. \ref{eqn:wave} is supplemented with an initial condition  ${E}(z,t=0)$. This condition describes a photon wave-packet immediately before the control field was applied. We have:
\begin{equation}
{E}(z,t=0)={F}_{s_0}(z,l_0),
\label{eqn:wave_init}
\end{equation}
This initial field is expanded over the set of the eigenfunctions of the common cavity with an effective length $l>l_0$. Within our approximations, a propagation of the initial field has an analytical solution:
\begin{equation}
{E}(z,t)=\underset{s}{\sum}C_s{F}_s(z,l)e^{i\omega_s t}
\label{eqn:wave_t}
\end{equation}
where the amplitudes $C_s=\langle F_s(z,l)|F_{s_0}(z,l_0)\rangle$ are the projections of the initial state on each of the eigenmodes of the common cavity and $\omega_s=k s c/n=\pi s c/(n l),\; s=1,2...$ 

\begin{figure}[t]
\includegraphics[width=0.95\columnwidth]{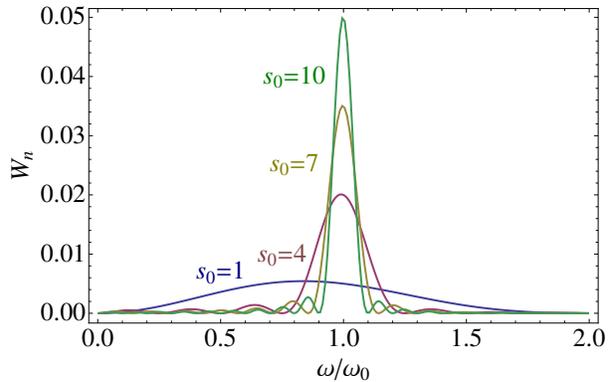}
\caption{\label{fig:pop} Normalized population of different eigenmodes $F_s(z,l)$ of the common resonator for different initially populated modes $s_0$ of the node cavity}
\end{figure}

The energy distribution over different eigenmodes $F_s(z,l)$ of the common cavity is given by:
\begin{equation}
W_s=\left|C_s\right|^2=\left[\frac{2 (-1)^{{s_0}}{s_0} \sin \left(\frac{\pi 
   {l_0} s}{l}\right)}{\left(\frac{1}{l}\right)^{3/2}
   \sqrt{\frac{1}{{l_0}}} \left(\pi  l^2 {s_0}^2-\pi 
   {l_0}^2 s^2\right)}\right]^2.
\label{eqn:wave_pop}
\end{equation}
Note that this distribution strongly depends on the ratio $l/l_0$ and the mode number of the populated mode in the isolated cavity $s_0$. Some examples of  normalized to have equal integral populations $W_s$ are shown in Fig. \ref{fig:pop} for different $s_0=1,...,10$ and for 
$l/l_0=200$. The distribution becomes narrower for larger values of $s_0$. A narrower distribution may be beneficial in material systems where it is difficult to avoid dispersion. On the other hand, coupling to higher modes decreases the coupling strength (i.e. the Rabi frequency in (\ref{eqn:Rabi})). Note the tradeoff between the coupling constant which decreases as $\sqrt{s_0}$, and the width of the distribution over eigenmodes, which narrows linearly with $s_0$. 

\begin{figure}[t]
\includegraphics[width=0.95\columnwidth]{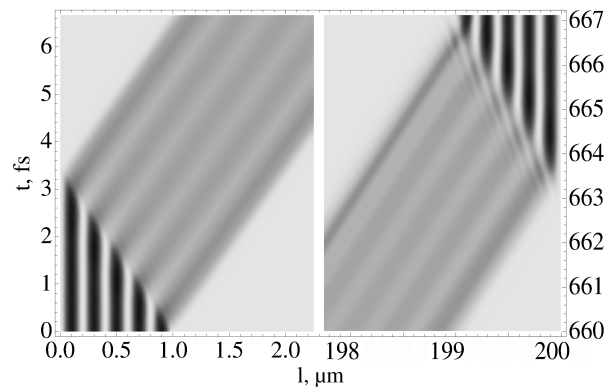}
\caption{\label{fig:two} Nearly-faithful restoration of field distribution $E(z,t=0)$ for an isolated cavity's mode $s_0=10$ after a propagation between the nodes in a dispersive waveguide (dispersion parameter $D=10$ ps$/($nm$\cdot$km$)$).}	
\end{figure}
\begin{figure}[t]
\includegraphics[width=0.95\columnwidth]{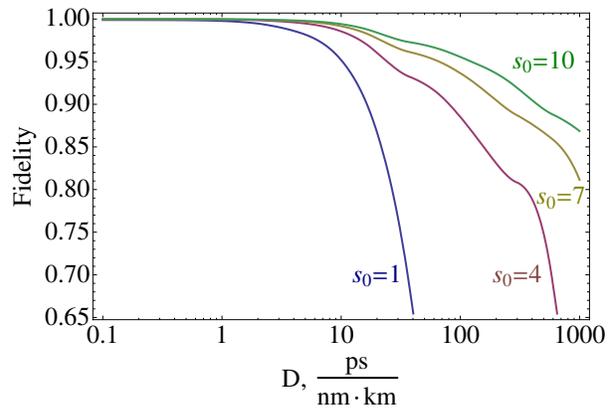}
\caption{\label{fig:fidelity} Fidelity for a round-trip photon wave-packet transfer (from one node to another and back) as a function of the group-velocity dispersion parameter $D$ for different isolated cavity modes $s_0$.}	
\end{figure}

We now include dispersion by introducing  $\omega_s=(\pi c/n) ((s/l)+(c D/n^2) (1-s/s_r)^2)$ into eq. (\ref{eqn:wave_t}), where $D$ is a group-velocity dispersion parameter and $s_r=s_0 l/l_0$ is the mode number of the resonant mode of the common cavity.  Because we consider eigenmodes of a node resonator, waveguide dispersion does not change the standing wave field shape, i.e. when the field $\lambda_2$ is off.
Figure \ref{fig:two} illustrates how the wave packet with $\lambda_1$ propagates during the photon exchange between the two nodes, where the stop band in-between the nodes is removed via a classical control optical field $\lambda_2$ with dispersion $D=10$ ps$/($nm$\cdot$km$)$. It is evident from Figure \ref{fig:two} that at the end of the transfer, the wavepacket has traveled to the target node with nearly unchanged amplitude and phase profiles. Therefore, upon propagation, a photon can readily interact with an ``atom'' in the target node once the control field is switched off.

Our subsequent simulation illustrates the effect of dispersion on fidelity $F$ of node-to-node photon state transfer. Because we use normalized eigenfunctions as initial conditions for (\ref{eqn:wave}), $F=Re\left[\overset{l_0}{\underset{0}{\int}}E(z,t=2l/(c/n))^{*}E(z,t=0)dx\right]$. We see that the higher the mode of an isolated cavity $s_0$ is used; the higher dispersion coefficients can be tolerated, Fig. \ref{fig:fidelity}. In particular, for a node-to-node transfer described above and $s_0=1$ the calculated fidelity $F>0.99$ for $D<2$ ps$/($nm$\cdot$km$)$, while for $s_0=10$ dispersion coefficients up to $D=20$ ps$/($nm$\cdot$km$)$ yield $F\approx0.99$, a ten-fold reduction in dispersion spreading.  Fidelity monotonically decreases as the group velocity dispersion increases. Ripples are seen in Fig.  \ref{fig:fidelity} due to the interference effects. 


\section{Two-qubit gate protocols}

Using isolated node manipulations and the controlled photon exchange between nodes we have developed simple and scalable protocols for basic quantum gates: SWAP, CNOT, and phase rotation. 
We define the two basic operations supported by the framework of an actively-switched photonic network and required to build quantum gates. We assume that the two cubits are stored in the two adjacent isolated cavities, labeled 1 and 2. First, a photon exchange between cavities through a common cavity is denoted by an exchange operator:
\begin{equation}
\hat{\leftrightarrow}|a_1\rangle_1|n_1\rangle_1|a_2\rangle_2|n_2\rangle_2=|a_1\rangle_1|n_2\rangle_1|a_2\rangle_2|n_1\rangle_2
\label{eqn:exch}
\end{equation}
where $|a\rangle_i$ is an atomic state and $|n\rangle_i$ is a photon number state in the $i^{th}$ cavity. Note that this procedure only exchanges the photonic part of the two bipartite cubits, leaving the node excitation part intact. 
\begin{table}[!h]
\centering
\caption{\bf Protocols for SWAP and CNOT operations}
{\footnotesize \def\arraystretch{1.1}\tabcolsep=0pt
\rowcolors{0.9}{}{gray!35}
\begin{tabular}{|c|c|c|c|c|}
\hline
\multicolumn{5}{|c|}{SWAP} \\
  \hline
  
$\hat{Op.}$ & $|g\rangle_1|1\rangle_1|g\rangle_2|1\rangle_2$ & $|g\rangle_1|1\rangle_1|e\rangle_2|0\rangle_2$ & $|e\rangle_1|0\rangle_1|g\rangle_2|1\rangle_2$ & $|e\rangle_1|0\rangle_1|e\rangle_2|0\rangle_2$ \\
$\hat{\leftrightarrow}$ & $|g\rangle_1|1\rangle_1|g\rangle_2|1\rangle_2$ & $|g\rangle_1|0\rangle_1|e\rangle_2|1\rangle_2$ & $|e\rangle_1|1\rangle_1|g\rangle_2|0\rangle_2$ & $|e\rangle_1|0\rangle_1|e\rangle_2|0\rangle_2$ \\
$\hat{\pi}$ & $|e\rangle_1|0\rangle_1|e\rangle_2|0\rangle_2$ & $|g\rangle_1|0\rangle_1|g\rangle_2|2\rangle_2$ & $|g\rangle_1|2\rangle_1|g\rangle_2|0\rangle_2$ & $|g\rangle_1|1\rangle_1|g\rangle_2|1\rangle_2$ \\
$\hat{\leftrightarrow}$ & $|e\rangle_1|0\rangle_1|e\rangle_2|0\rangle_2$ & $|g\rangle_1|2\rangle_1|g\rangle_2|0\rangle_2$ & $|g\rangle_1|0\rangle_1|g\rangle_2|2\rangle_2$ & $|g\rangle_1|1\rangle_1|g\rangle_2|1\rangle_2$ \\
$\hat{\pi}$ & $|g\rangle_1|1\rangle_1|g\rangle_2|1\rangle_2$ & $|e\rangle_1|1\rangle_1|g\rangle_2|0\rangle_2$ & $|g\rangle_1|0\rangle_1|e\rangle_2|1\rangle_2$ & $|e\rangle_1|0\rangle_1|e\rangle_2|0\rangle_2$ \\
$\hat{\leftrightarrow}$ & $|g\rangle_1|1\rangle_1|g\rangle_2|1\rangle_2$ & $|e\rangle_1|0\rangle_1|g\rangle_2|1\rangle_2$ & $|g\rangle_1|1\rangle_1|e\rangle_2|0\rangle_2$ & $|e\rangle_1|0\rangle_1|e\rangle_2|0\rangle_2$ \rowcolors{row}{gray !20}{}\\
  \hline
\hline

\multicolumn{5}{|c|}{CNOT} \\
  \hline

$\hat{Op.}$ & $|g\rangle_1|1\rangle_1|g\rangle_2|1\rangle_2$ & $|g\rangle_1|1\rangle_1|e\rangle_2|0\rangle_2$ & $|e\rangle_1|0\rangle_1|g\rangle_2|1\rangle_2$ & $|e\rangle_1|0\rangle_1|e\rangle_2|0\rangle_2$ \\
$\hat{\leftrightarrow}$ & $|g\rangle_1|1\rangle_1|g\rangle_2|1\rangle_2$ & $|g\rangle_1|0\rangle_1|e\rangle_2|1\rangle_2$ & $|e\rangle_1|1\rangle_1|g\rangle_2|0\rangle_2$ & $|e\rangle_1|0\rangle_1|e\rangle_2|0\rangle_2$ \\
$\hat{\pi}$ & $|e\rangle_1|0\rangle_1|e\rangle_2|0\rangle_2$ & $|g\rangle_1|0\rangle_1|g\rangle_2|2\rangle_2$ & $|g\rangle_1|2\rangle_1|g\rangle_2|0\rangle_2$ & $|g\rangle_1|1\rangle_1|g\rangle_2|1\rangle_2$ \\
$\hat{\leftrightarrow}$ & $|e\rangle_1|0\rangle_1|e\rangle_2|0\rangle_2$ & $|g\rangle_1|2\rangle_1|g\rangle_2|0\rangle_2$ & $|g\rangle_1|0\rangle_1|g\rangle_2|2\rangle_2$ & $|g\rangle_1|1\rangle_1|g\rangle_2|1\rangle_2$ \\
$\hat{\pi}_1$ & $|g\rangle_1|1\rangle_1|e\rangle_2|0\rangle_2$ & $|e\rangle_1|1\rangle_1|g\rangle_2|0\rangle_2$ & $|g\rangle_1|0\rangle_1|g\rangle_2|2\rangle_2$ & $|e\rangle_1|0\rangle_1|g\rangle_2|1\rangle_2$ \\
$\hat{\leftrightarrow}$ & $|g\rangle_1|0\rangle_1|e\rangle_2|1\rangle_2$ & $|e\rangle_1|0\rangle_1|g\rangle_2|1\rangle_2$ & $|g\rangle_1|2\rangle_1|g\rangle_2|0\rangle_2$ & $|e\rangle_1|1\rangle_1|g\rangle_2|0\rangle_2$ \\
$\hat{\pi}_2$ & $|g\rangle_1|0\rangle_1|g\rangle_2|2\rangle_2$ & $|e\rangle_1|0\rangle_1|e\rangle_2|0\rangle_2$ & $|g\rangle_1|2\rangle_1|g\rangle_2|0\rangle_2$ & $|e\rangle_1|1\rangle_1|g\rangle_2|0\rangle_2$ \\
$\hat{\leftrightarrow}$ & $|g\rangle_1|2\rangle_1|g\rangle_2|0\rangle_2$ & $|e\rangle_1|0\rangle_1|e\rangle_2|0\rangle_2$ & $|g\rangle_1|0\rangle_1|g\rangle_2|2\rangle_2$ & $|e\rangle_1|0\rangle_1|g\rangle_2|1\rangle_2$ \\
$\hat{\pi}_1$ & $|e\rangle_1|1\rangle_1|g\rangle_2|0\rangle_2$ & $|g\rangle_1|1\rangle_1|e\rangle_2|0\rangle_2$ & $|g\rangle_1|0\rangle_1|g\rangle_2|2\rangle_2$ & $|g\rangle_1|1\rangle_1|g\rangle_2|1\rangle_2$ \\
$\hat{\leftrightarrow}$ & $|e\rangle_1|0\rangle_1|g\rangle_2|1\rangle_2$ & $|g\rangle_1|0\rangle_1|e\rangle_2|1\rangle_2$ & $|g\rangle_1|2\rangle_1|g\rangle_2|0\rangle_2$ & $|g\rangle_1|1\rangle_1|g\rangle_2|1\rangle_2$ \\
$\hat{\pi}_2$ & $|e\rangle_1|0\rangle_1|e\rangle_2|0\rangle_2$ & $|g\rangle_1|0\rangle_1|g\rangle_2|2\rangle_2$ & $|g\rangle_1|2\rangle_1|g\rangle_2|0\rangle_2$ & $|g\rangle_1|1\rangle_1|e\rangle_2|0\rangle_2$ \\
$\hat{\leftrightarrow}$ & $|e\rangle_1|0\rangle_1|e\rangle_2|0\rangle_2$ & $|g\rangle_1|2\rangle_1|g\rangle_2|0\rangle_2$ & $|g\rangle_1|0\rangle_1|g\rangle_2|2\rangle_2$ & $|g\rangle_1|0\rangle_1|e\rangle_2|1\rangle_2$ \\
$\hat{\pi}_1$ & $|g\rangle_1|1\rangle_1|e\rangle_2|0\rangle_2$ & $|e\rangle_1|1\rangle_1|g\rangle_2|0\rangle_2$ & $|g\rangle_1|0\rangle_1|g\rangle_2|2\rangle_2$ & $|g\rangle_1|0\rangle_1|e\rangle_2|1\rangle_2$ \\
$\hat{\leftrightarrow}$ & $|g\rangle_1|0\rangle_1|e\rangle_2|1\rangle_2$ & $|e\rangle_1|0\rangle_1|g\rangle_2|1\rangle_2$ & $|g\rangle_1|2\rangle_1|g\rangle_2|0\rangle_2$ & $|g\rangle_1|1\rangle_1|e\rangle_2|0\rangle_2$ \\
$\hat{\pi}_2$ & $|g\rangle_1|0\rangle_1|g\rangle_2|2\rangle_2$ & $|e\rangle_1|0\rangle_1|e\rangle_2|0\rangle_2$ & $|g\rangle_1|2\rangle_1|g\rangle_2|0\rangle_2$ & $|g\rangle_1|1\rangle_1|g\rangle_2|1\rangle_2$ \\
$\hat{\pi}$ & $|g\rangle_1|0\rangle_1|e\rangle_2|1\rangle_2$ & $|g\rangle_1|1\rangle_1|g\rangle_2|1\rangle_2$ & $|e\rangle_1|1\rangle_1|g\rangle_2|0\rangle_2$ & $|e\rangle_1|0\rangle_1|e\rangle_2|0\rangle_2$ \\
$\hat{\leftrightarrow}$ & $|g\rangle_1|1\rangle_1|e\rangle_2|0\rangle_2$ & $|g\rangle_1|1\rangle_1|g\rangle_2|1\rangle_2$ & $|e\rangle_1|0\rangle_1|g\rangle_2|1\rangle_2$ & $|e\rangle_1|0\rangle_1|e\rangle_2|0\rangle_2$ \\
  \hline

\end{tabular}
} 
  \label{tab:protocols}
\end{table}
Second, an inversion operator is naturally provided by Rabi oscillations in the nodes. When $2\pi$-pulses are applied to the selected group of nodes to freeze their evolution, the other nodes undergo an inversion:
\begin{eqnarray}
\hat{\pi}|g\rangle|n\rangle=|e\rangle|n-1\rangle \\
\hat{\pi}|e\rangle|n\rangle=|g\rangle|n+1\rangle,
\label{eqn:pi}
\end{eqnarray}
where $\hat{\pi}$ - is a delay equal to one half of Rabi oscillations period during the evolution of a node. In the Table, $\hat{\pi}$ refers to an inversion on both the nodes, $\hat{\pi}_i$, where $i=1,2$ denotes an inversion of just the first (second) node, while the other one is not inverted.

The above operators are sufficient to implement SWAP and CNOT gates. The associated protocols are shown in Table \ref{tab:protocols}.

\section{Conclusions}

We have introduced a method of implementing scalable light-controlled gates for quantum information processing, based on N-level systems (nodes) exchanging photons via optically-controlled nonlinear Bragg waveguides. This method for all-optical switching relies on cross-phase modulation that removes the effective Bragg resonance and creates transparency so that the control field can be detuned very far from any of the node's resonances. Because the control field $\lambda_2$ is a classical field, multiple pairs of nodes may be controlled simultaneously. Thus, quantum information processing can be made massively parallel. This design can be implemented on a chip with the 
currently existing technology. A particular experimental implementation of this method is described in the Supplementary material section. Note, that our proposal only requires a fast (femtosecond scale) classical switch. The search for fast, high-contrast optically controlled switches is a very active field of research, \cite{Midwinter,Sivan,Zang,Masaaki,LIDG}. Therefore one should expect alternative experimental realizations of the proposed protocol.

While, in our manuscript, we assumed that the control field is applied perpendicular to the waveguide, other configurations are also possible. To implement even more exotic, multi-body based algorithms \cite{Greiner}, nodes can be arranged in one-, two-, or three-dimensional structures. This opens the way to implement topologically protected and/or massively parallel quantum interactions, commonly studied with cold atoms in optical lattices, on a solid state chip.

\section*{SUPPLEMENTARY MATERIAL}

We discussed the general idea and introduced protocols for a quantum circuit based on actively switched optical channels. Because a broad range of materials and geometries can be employed to implement these circuits \cite{Vukovic,Yoshiki,Yuce,Angelis} we presented general design features for such channels. Here we offer a detailed study of one possible implementation of our protocol, based on a lithium niobate waveguide. The purpose of this study is twofold. First, it provides guidance for an enabling experiment with Bragg switches. Second, it introduces certain technological enhancements that are aimed at scalability.

Lithium niobate is commonly used in optical devices and is relatively well understood. We consider it a good candidate for an experimental realization because it has one of the largest Kerr nonlinearity constants $n_2=83.3 10^{-16}$ cm$^2/$W \cite{Nikogosian}. All necessary fabrication techniques required for realization of our proposal, particularly the making of the waveguides \cite{Pierno}, the permanent modulation of the refraction index \cite{Paipulas,Horn}, the making of the dynamic gratings with a static electrical field \cite{Jin} and placing the conductive nanostructures \cite{AuLiNbO3} were demonstrated. The making of the dynamic gratings with a light field \cite{Schneider} as well as field enhancement with periodic conductive nanostructures \cite{Gogol} were demonstrated in other materials, but are compatible with lithium niobate.

The overall geometry of the structure and the mode profile is shown in Fig. 1 of the manuscript. The waveguide can be implemented in lithium niobate with an average diameter of $\approx500$ nm on a lower refractive index (RI) substrate with.  Periodic variation of the RI of the waveguide can be implemented statically in several different ways or dynamically induced by a control field. An RI contrast is chosen based on the nonlinear constant, available control field power and the required bandgap width. 

\begin{figure}
\center\includegraphics[width=0.9\columnwidth]{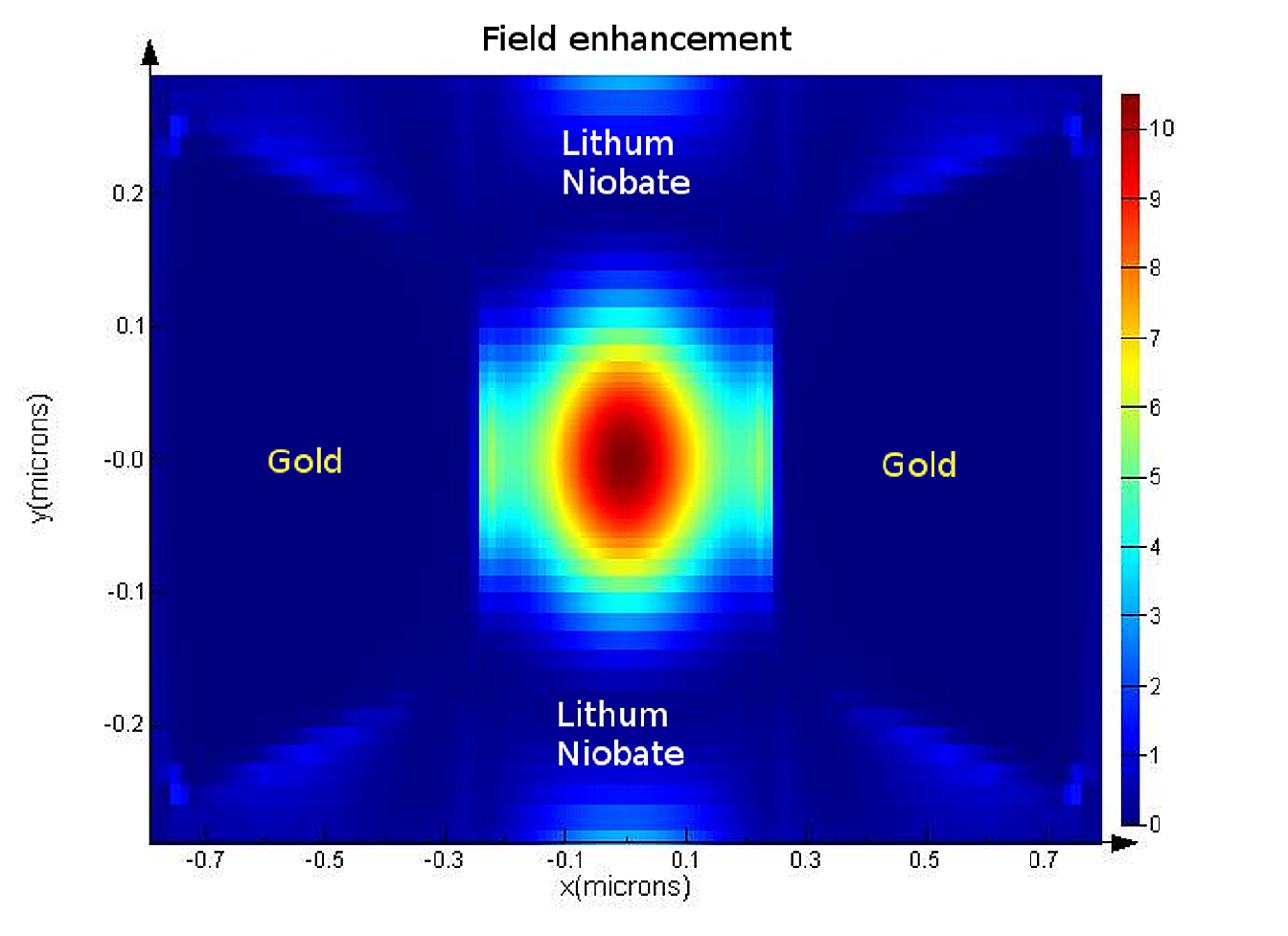}
\caption{Field intensity enhancement in $\mathrm{LiNbO_3}$ waveguide due to plasmonic resonance of gold nanoantenna (dark blue trapezoidal regions) with air cladding}
\label{fig:five}
\end{figure}

To implement the proposed switching mechanism in a lithium niobate waveguide, a low RI contrast should be used, such that a control field of $\approx 10^{10}\ldots10^{11}$ W/cm$^2$ can erase a static Bragg grating (BG). This intensity corresponds to a refractive index change due to a Kerr cross-phase modulation of $\approx10^{-3}$. To make a high Q cavity with such a low RI contrast, the length of the grating structure should be quite long: $\approx3000$ periods. Notice that the bandgap width will also decrease ($\delta\lambda\approx0.5$ nm
, thus a higher longitudinal cavity mode should be coupled to the quantum node to provide adequately high reflectivity ($s_0\approx100$). To erase a BG structure, the control field should have a spatial pattern that corresponds to the periodic BG structure. 
An array of plasmonic antennas along the waveguide could be used to enhance the local optical field and provide modulation \cite{Bruck,Gogol}, thus reducing the power required and simplifying the preparation of the control beams.
The effect of placing the antennas is shown in Fig. \ref{fig:five}, where a numerical calculation was performed with an FDTD method for a control field with a wavelength $\lambda=1$ $\mathrm{\mu m}$. As a result, a simple flat-top (and unmodulated) control field can be applied as a control, further aiding scalability.  In addition, such an array significantly decreases the power requirements for the control field. 
Eliminating the BG altogether is advantageous because it reduces undesired dispersion effects. We have modeled the propagation of a wave packet through the waveguide and it is evident that dispersion in this system does not significantly distort the wavepacket, yielding a fidelity above $0.99$ for the 10$^{th}$ mode (c.f. Fig. 4 of the manuscript). As seen in that figure, the fidelity improves further at even higher mode numbers. Placing an array of conducting particles in the vicinity of the waveguide does not significantly increase the propagation loss because the $\lambda_1$ field is far off-resonant for gold nanoantennas. 

Assuming that the control field is sufficiently detuned from the transition frequency used in the 2-level nodes ($\Delta\lambda>100$ nm), a qubit dipole moment of $\approx ea_0$ (where $a_0$ is Bohr radius) and a resonant energy of $\approx1$ eV \cite{Delone}, the resulting Rabi oscillations of the qubit due to the control field 
yield an amplitude that is less than $10^{-3}$, i.e. negligible on the time scale in question.  Minor drawbacks of using shallow index contrast vs. high index contrast are the relatively long distances between nodes, longer times for gate operations, and longer Rabi periods for the coupling between nodes and isolated cavities (up to 10 times in comparison to the lowest longitudinal mode attainable with high index contrast). 

On the other hand, high RI contrast, for instance $\delta n=0.04$, would result in a wide band-gap ($\delta\lambda\approx 10$ nm, providing finesse $f\approx10^6$ for $10^{th}$ field mode) with only $200$ periods. When a large index contrast is employed, an optically controlled bandgap shift, rather than a BG erasure, should be used to lower switching power requirements. Remarkably,  a required control field pulse peak power of $\approx 720$ W and a duration of $\approx 100$ ps is sufficient to achieve bandgap shifts of $\approx 0.6$ nm in barium fluoride \cite{shift, Zang}. Due to a larger Kerr coefficient, this power is reduced  more than 30-fold  in lithium niobate. To reduce the role of strong dispersion as the light propagates in the medium with an optically detuned bandgap, a higher-order longitudinal isolated cavity mode should be employed (see Fig. 2 of the manuscript), and/or dispersion compensation techniques should be applied. Dispersion engineering can be achieved e.g. through advanced 2D patterning of lithium niobate \cite{disp}.

In conclusion, we discussed general design rules using one possible experimental implementation of the platform as an example. In addition an array of plasmonic antennas along the waveguide enhances the local optical field and provides modulation, thus aids scalability. 

\section*{ACKNOWLEDGEMENTS}

Authors thank Mohammad Hafezi for fruitful discussions and critical comments. SVP thanks Garnett Bryant for encouraging remarks. 
IAB and OVT acknowledge a partial support by Russian Foundation for Basic Research grant No 14-02-00389\_a.

\section*{References}
\bibliography{biblio}

\providecommand{\newblock}{}
\begin{thebibliography}{10}
\expandafter\ifx\csname url\endcsname\relax
  \def\url#1{{\tt #1}}\fi
\expandafter\ifx\csname urlprefix\endcsname\relax\def\urlprefix{URL }\fi
\providecommand{\eprint}[2][]{\url{#2}}

\bibitem{Ekert}
Ekert A and Jozsa R 1998 {\em Philosophical Transactions of the Royal Society
  of London A: Mathematical, Physical and Engineering Sciences\/} {\bf 356}
  1769--1782 ISSN 1364-503X

\bibitem{Shor}
Shor P 2004 {\em Quantum Information Processing\/} {\bf 3} 5--13 ISSN 1570-0755
  \urlprefix\url{http://dx.doi.org/10.1007/s11128-004-3878-2}

\bibitem{DiVi1}
DiVincenzo D~P 1995 {\em Science\/} {\bf 270} 255--261

\bibitem{Ions}
Home J~P, Hanneke D, Jost J~D, Amini J~M, Leibfried D and Wineland D~J 2009
  {\em Science\/} {\bf 325} 1227--1230 ISSN 0036-8075

\bibitem{QDot}
Gazzano O and Solomon G~S 2016 {\em J. Opt. Soc. Am. B\/} {\bf 33} C160--C175
  \urlprefix\url{http://josab.osa.org/abstract.cfm?URI=josab-33-7-C160}

\bibitem{RAtoms}
Saffman M, Walker T~G and M$\ddot{\mathrm{o}}$lmer K 2010 {\em Rev. Mod.
  Phys.\/} {\bf 82} 2313--2363
  \urlprefix\url{http://link.aps.org/doi/10.1103/RevModPhys.82.2313}

\bibitem{Atoms}
Monroe C 2002 {\em Nature\/} {\bf 416} 238--246 ISSN 0028-0836
  \urlprefix\url{http://dx.doi.org/10.1038/416238a}

\bibitem{Monroe}
Kielpinski D, Monroe C and Wineland D~J 2002 {\em Nature\/} {\bf 417} 709--711
  ISSN 0028-0836

\bibitem{MIons}
Mavadia S, Goodwin J~F, Stutter G, Bharadia S, Crick D~R, Segal D~M and
  Thompson R~C 2013 {\em Nat Commun\/} {\bf 4}
  \urlprefix\url{http://dx.doi.org/10.1038/ncomms3571 10.1038/ncomms3571}

\bibitem{spin}
Sahling S, Remenyi G, Paulsen C, Monceau P, Saligrama V, Marin C, Revcolevschi
  A, Regnault L~P, Raymond S and Lorenzo J~E 2015 {\em Nat Phys\/} {\bf 11}
  255--260 ISSN 1745-2473 article
  \urlprefix\url{http://dx.doi.org/10.1038/nphys3186}

\bibitem{NVC}
Dutt M~V~G, Childress L, Jiang L, Togan E, Maze J, Jelezko F, Zibrov A~S,
  Hemmer P~R and Lukin M~D 2007 {\em Science\/} {\bf 316} 1312--1316
  (\textit{Preprint}
  \eprint{http://www.sciencemag.org/content/316/5829/1312.full.pdf})
  \urlprefix\url{http://www.sciencemag.org/content/316/5829/1312.abstract}

\bibitem{Nozaki}
Nozaki K, Tanabe T, Shinya A, Matsuo S, Sato T, Taniyama H and Notomi M 2010
  {\em Nature Photonics\/} {\bf 4} 477--483 ISSN 1749-4885
  \urlprefix\url{http://www.nature.com/doifinder/10.1038/nphoton.2010.89}

\bibitem{Schneider}
Schneider T, Wolfframm D, Mitzner R and Reif J 1999 {\em Applied Physics B\/}
  {\bf 68} 749--751 ISSN 0946-2171
  \urlprefix\url{http://dx.doi.org/10.1007/s003400050698}

\bibitem{Jin}
Jin W, Chiang K~S and Liu Q 2008 {\em Opt. Express\/} {\bf 16} 20409--20417
  \urlprefix\url{http://www.opticsexpress.org/abstract.cfm?URI=oe-16-25-20409}

\bibitem{Monroe1}
Luo L, Hayes D, Manning T, Matsukevich D, Maunz P, Olmschenk S, Sterk J and
  Monroe C 2009 {\em Fortschritte der Physik\/} {\bf 57} 1133--1152 ISSN
  1521-3978 \urlprefix\url{http://dx.doi.org/10.1002/prop.200900093}

\bibitem{Pelton}
Pelton M, Santori C, Vuckovic J, Zhang B~Y, Solomon G~S, Plant J and Yamamoto Y
  2002 {\em Phys. Rev. Lett.\/} {\bf 89} 233602 ISSN 0031-9007

\bibitem{StrongCouple}
Reithmaier J~P, Sek G, Loffler, Hofmann C, Kuhn S, Reitzenstein S, Keldysh L~V,
  Kulakovskii V~D, Reinecke T~L and Forchel A 2004 {\em Nature\/} {\bf 432}
  197--200
  \urlprefix\url{http://www.nature.com/nature/journal/v432/n7014/abs/nature02969.html}

\bibitem{StrongCouple2}
Yoshie T, Scherer A, Hendrickson J, Khitrova G, Gibbs H~M, Rupper G, Ell C,
  Shchekin O~B and Deppe D~G 2004 {\em Nature\/} {\bf 432} 200--203
  \urlprefix\url{http://www.nature.com/nature/journal/v432/n7014/abs/nature02969.html}

\bibitem{Bose}
Bose R, Cai T, Choudhury K~R, Solomon G~S and Waks E 2014 {\em Nature
  Photonics\/} {\bf 8} 858--864 ISSN 1749-4885
  \urlprefix\url{http://www.nature.com/doifinder/10.1038/nphoton.2014.224}

\bibitem{Pierno}
Pierno L, Dispenza M, Secchi A, Fiorello A and Foglietti V 2008 {\em Journal of
  Optics A: Pure and Applied Optics\/} {\bf 10} 064017
  \urlprefix\url{http://stacks.iop.org/1464-4258/10/i=6/a=064017}

\bibitem{EberlyBook}
Allen L and Eberly J~H 1987 {\em Optical Resonance and Two-Level Atoms\/}
  (Dover: New York)

\bibitem{APass}
Wei Y~J, He Y~M, Chen M~C, Hu Y~N, He Y, Wu D, Schneider C, Kamp M, Höfling S,
  Lu C~Y and Pan J~W 2014 {\em Nano Letters\/} {\bf 14} 6515--6519 pMID:
  25357153 (\textit{Preprint} \eprint{http://dx.doi.org/10.1021/nl503081n})
  \urlprefix\url{http://dx.doi.org/10.1021/nl503081n}

\bibitem{PhBlockade}
Birnbaum K~M, Boca A, Miller R, Boozer A~D, Northup T~E and Kimble H~J 2005
  {\em Nature\/} {\bf 436} 87--90 ISSN 0028-0836
  \urlprefix\url{http://dx.doi.org/10.1038/nature03804}

\bibitem{Midwinter}
Midwinter J, Liao P and Kelley P 2012 {\em Photonics in Switching\/} Quantum
  Electronics--Principles and Applications (Elsevier Science) ISBN
  9780080924748 \urlprefix\url{https://books.google.com/books?id=WYBYJpET-dwC}

\bibitem{Sivan}
Sivan Y, Ctistis G, Y\"{u}ce E and Mosk A~P 2015 {\em Opt. Express\/} {\bf 23}
  16416--16428
  \urlprefix\url{http://www.opticsexpress.org/abstract.cfm?URI=oe-23-12-16416}

\bibitem{Zang}
Zang Z and Zhang Y 2012 {\em Appl. Opt.\/} {\bf 51} 3424--3430
  \urlprefix\url{http://ao.osa.org/abstract.cfm?URI=ao-51-16-3424}

\bibitem{Masaaki}
Tada K 2005 {\em Photonics Based on Wavelength Integration and Manipulation\/}
  IPAP books (IPAP) ISBN 9784900526198
  \urlprefix\url{https://books.google.com/books?id=tTlzAQAACAAJ}

\bibitem{LIDG}
Eichler H, G{\"u}nter P and Pohl D 1986 {\em Laser-induced Dynamic Gratings\/}
  Exploration of the Deep Continental Crust (Springer-Verlag) ISBN
  9780387158754 \urlprefix\url{https://books.google.com/books?id=N3lrQgAACAAJ}

\bibitem{Greiner}
Greiner M, Mandel O, Esslinger T, Hansch T~W and Bloch I 2002 {\em Nature\/}
  {\bf 415} 39--44 ISSN 0028-0836
  \urlprefix\url{http://dx.doi.org/10.1038/415039a}

\bibitem{Vukovic}
Vukovic N, Healy N, Suhailin F~H, Mehta P, Day T~D, Badding J~V and Peacock A~C
  2013 {\em Sci Rep\/} {\bf 3} 2885 ISSN 2045-2322 24097126[pmid]
  \urlprefix\url{http://www.ncbi.nlm.nih.gov/pmc/articles/PMC3791441/}

\bibitem{Yoshiki}
Yoshiki W and Tanabe T 2014 {\em Opt. Express\/} {\bf 22} 24332--24341
  \urlprefix\url{http://www.opticsexpress.org/abstract.cfm?URI=oe-22-20-24332}

\bibitem{Yuce}
Thyrrestrup H, Yüce E, Ctistis G, Claudon J, Vos W~L and Gérard J~M 2014 {\em
  Applied Physics Letters\/} {\bf 105} 111115

\bibitem{Angelis}
De~Angelis C, Modotto D, Locatelli A and Wabnitz S 2015 {\em Optical Guided
  Wave Switching\/} (Cham: Springer International Publishing) pp 71--104 ISBN
  978-3-319-14992-9

\bibitem{Nikogosian}
Nikogosian D~N 1997 {\em Properties of Optical and Laser-Related Materials: A
  Handbook\/} (Wiley Interscience)

\bibitem{Paipulas}
Paipulas D, Malinauskas M, Smilgevičius V and Sirutkaitis V 2011 Permanent
  volume bragg grating fabrication in pure lithium niobate crystal using direct
  laser writing technique {\em Lasers and Electro-Optics Europe (CLEO
  EUROPE/EQEC), 2011 Conference on and 12th European Quantum Electronics
  Conference\/} pp 1--1 ISSN Pending

\bibitem{Horn}
Horn W, Kroesen S, Herrmann J, Imbrock J and Denz C 2012 {\em Opt. Express\/}
  {\bf 20} 26922--26928
  \urlprefix\url{http://www.opticsexpress.org/abstract.cfm?URI=oe-20-24-26922}

\bibitem{AuLiNbO3}
Green T~A 2007 {\em Gold Bulletin\/} {\bf 40} 105--114 ISSN 2190-7579
  \urlprefix\url{http://dx.doi.org/10.1007/BF03215566}

\bibitem{Gogol}
F\'{e}vrier M, Gogol P, Barbillon G, Aassime A, M\'{e}gy R, Bartenlian B,
  Lourtioz J~M and Dagens B 2012 {\em Opt. Express\/} {\bf 20} 17402--17409
  \urlprefix\url{http://www.opticsexpress.org/abstract.cfm?URI=oe-20-16-17402}

\bibitem{Bruck}
Bruck R and Muskens O~L 2013 {\em Opt. Express\/} {\bf 21} 27652--27661
  \urlprefix\url{http://www.opticsexpress.org/abstract.cfm?URI=oe-21-23-27652}

\bibitem{Delone}
Delone N and Kra{\u\i}nov V 2000 {\em Multiphoton Processes in Atoms: Second
  Edition\/} Atoms and plasmas (Springer) ISBN 9783540646150
  \urlprefix\url{https://books.google.com/books?id=EJhSVHGe5DMC}

\bibitem{shift}
Yokota H, Kobayashi M, Mineo H, Kagawa N, Kanbe H and Sasaki Y 2008 {\em Optics
  Communications\/} {\bf 281} 4893 -- 4898 ISSN 0030-4018
  \urlprefix\url{http://www.sciencedirect.com/science/article/pii/S003040180800597X}

\bibitem{disp}
Broderick N~G~R, Ross G~W, Offerhaus H~L, Richardson D~J and Hanna D~C 2000
  {\em Phys. Rev. Lett.\/} {\bf 84}(19) 4345--4348
  \urlprefix\url{http://link.aps.org/doi/10.1103/PhysRevLett.84.4345}

\end{thebibliography}

\end{document}